\documentclass[twocolumn,twoside,slac_two]{revtex4}
\usepackage{graphicx}
\usepackage{fancyhdr}
\begin{document}

\title{VERITAS Observations of High-Frequency-Peaked BL Lac Objects}

%

\author{M. Orr}
\affiliation{Iowa State University, 12 Physics Hall, Ames, Iowa, 50011, United States}
\author{for the VERITAS Collaboration}

\begin{abstract}
Here we present highlights from VERITAS observations of high-frequency-peaked BL Lac objects (HBLs). We discuss the key science motivations for observing these sources -- including performing multiwavelength campaigns critical to understanding the emission mechanisms at work in HBLs, constraining the intensity and spectra shape of the extragalactic background light, and placing limits on the strength of the intergalactic magnetic field.
\end{abstract}

\maketitle

\thispagestyle{fancy}


\section{VERITAS Observatory}
The VERITAS observatory is an array of four 12-meter imaging atmospheric-Cherenkov telescopes (IACTs) located at the Fred Lawrence Whipple Observatory (FLWO) in southern Arizona \citep{Holder:2008}. Each reflector comprises 350 hexagonal mirrors following the Davies-Cotton design. The focal plane instrument consists of 499 photomultiplier tubes covering a 3.5$^\circ$ field of view. Its large collection area ($\sim \negthinspace 105 \, \mathrm{m}^2$), in conjunction with the stereoscopic imaging of air showers, allows VERITAS to detect very-high-energy (VHE) gamma rays between energies of $100\,$GeV and $30\,$TeV with an energy resolution of $\sim \negthinspace 15 \negthinspace - \negthinspace 20$\% and an angular resolution of $\sim \negthinspace 0.1^\circ$. A source with a flux 1\% that of the Crab Nebula can be detected by VERITAS in $\sim \negthinspace 25\,$hours with a statistical significance of 5 standard deviations ($\sigma$).\footnote{This refers to the instrument snesitivity prior to the VERITAS camera upgrade in 2012.}

During the bright-moonlight period in November 2011, the VERITAS telescope-level trigger systems were upgraded.  This was part of a larger upgrade of the VERITAS array which included the replacement of all camera pixels with high quantum efficiency photomultiplier tubes during the summer of 2012. 

A significant component of the VERITAS science and observing program involves the study of active galactic nuclei (AGNs).  In these proceedings we outline the observations of a specific class of AGNs known as high-frequency-peaked BL Lac objects (HBLs).  The VERITAS long-term blazar science and observing program consists, in part, of obtaining $\sim \negthinspace 200\,$ hours of total exposure on several key HBLs.  In Section \ref{sec:hbls} we briefly outline the characteristics of AGNs, and HBLs in particular.  Section \ref{sec:science} outlines the broad-ranging science motivations for studying these objects.  Concluding remarks are given in Section \ref{sec:conclusion}.

\section{High-Frequency-Peaked BL Lacs}
\label{sec:hbls}
\begin{figure*}[t]
\centering
\includegraphics[width=155mm]{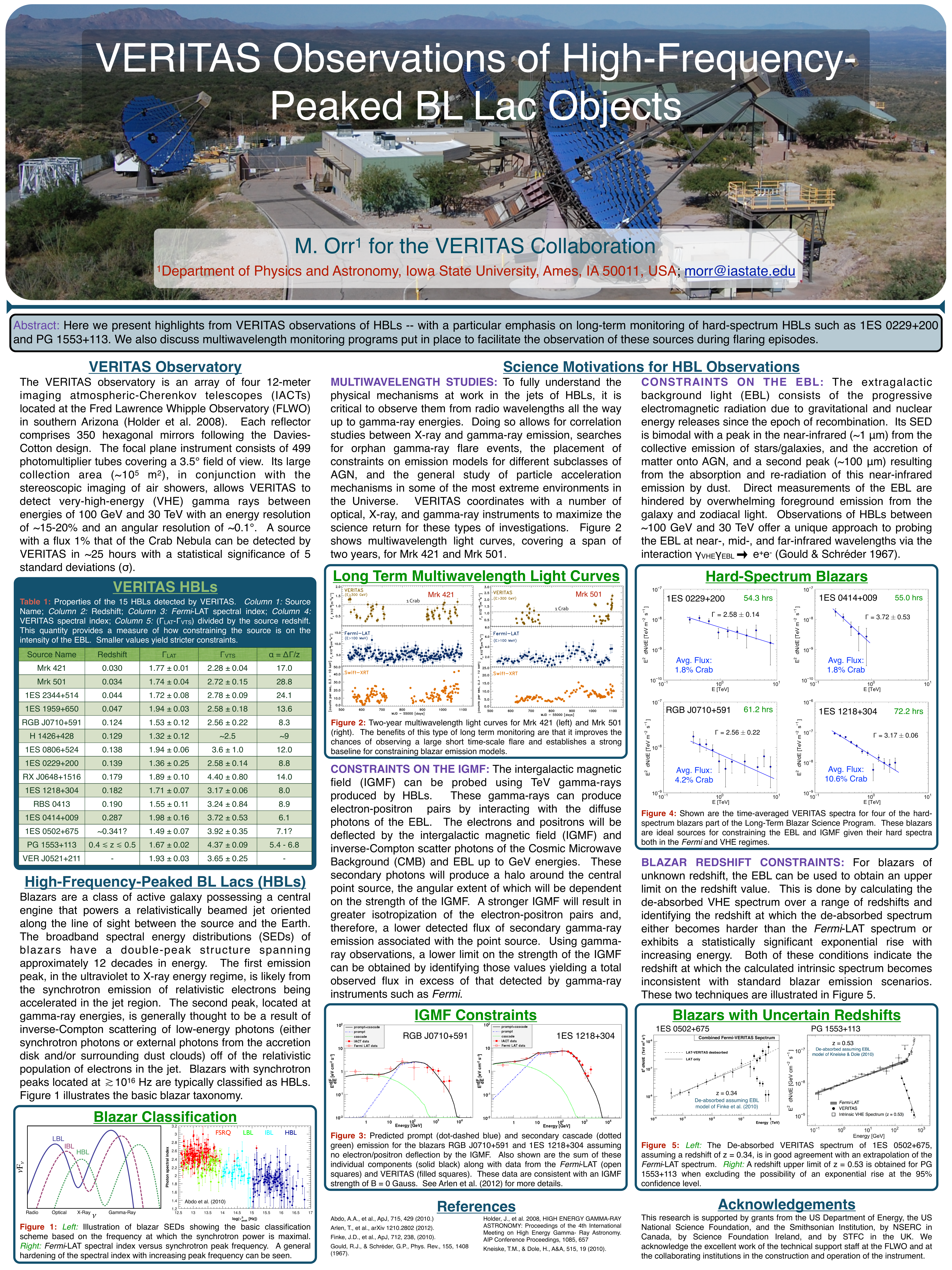}
\caption{\textit{Left:} Illustration of blazar SEDs showing the basic classification scheme based on the frequency at which the synchrotron power is maximal.  \textit{Right:} Fermi-LAT spectral index versus synchrotron peak frequency.  A general hardening of the spectral index with increasing peak frequency can be seen. (From \citet{Abdo:2010}.)}
\label{fig:hbls}
\end{figure*}

Blazars are a class of active galaxy possessing a central engine that powers a relativistically beamed jet oriented along the line of sight between the source and the Earth.  The broadband spectral energy distributions (SEDs) of blazars have a double-peak structure spanning approximately 12 decades in energy.  The first emission peak, in the ultraviolet to X-ray energy regime, is likely from the synchrotron emission of relativistic electrons being accelerated in the jet region.  The second peak, located at gamma-ray energies, is generally thought to be a result of inverse-Compton scattering of low-energy photons (either synchrotron photons or external photons from the accretion disk and/or surrounding dust clouds) off the relativistic population of electrons in the jet.  Blazars with synchrotron peaks located at $\geq \negthinspace 10^{16}$Hz are typically classified as HBLs \citep{Abdo:2010}.\footnote{\citet{Abdo:2010}, in fact, use the terminology high-synchrotron-peaked BL Lac object (HSP) rather than HBL.  In the context of this paper, the two naming schemese are considered interchangeable.} Figure \ref{fig:hbls} illustrates the basic blazar taxonomy.

The 15 VERITAS-detected HBLs are listed in Table \ref{tab:hbls}.  The source redshift, \textit{Fermi} and VERITAS spectral indices, and \textit{Fermi}/VERITAS spectral break divided by the source redshift are given.  

\begin{table*}[t]
\begin{center}
\caption{Properties of the 15 HBLs detected by VERITAS.  \textit{Column 1:} Source Name; \textit{Column 2:} Redshift; \textit{Column 3:} Fermi-LAT spectral index; \textit{Column 4:} VERITAS spectral index; \textit{Column 5:} $(\Gamma_\mathrm{VTS} - \Gamma_\mathrm{LAT})$ divided by the source redshift.  This quantity provides a measure of how constraining the source is on the intensity of the EBL.  Smaller values yield stricter constraints.}
\begin{tabular}{| c | c | c | c | c |}
\hline
Source Name & Redshift & $\Gamma_\mathrm{LAT}$ & $\Gamma_\mathrm{VTS}$ & $\alpha = \Delta\Gamma / z$ \\
\hline
Mrk 421 & 0.030 &	$1.77 \pm 0.01$ &	$2.28 \pm 0.04$ & 17.0 \\
Mrk 501 & 0.034 &	$1.74 \pm 0.04$ &	$2.72 \pm 0.15$ &	28.8 \\
1ES 2344+514 & 0.044	& $1.72 \pm 0.08$ & $2.78 \pm 0.09$ & 24.1 \\
1ES 1959+650 & 0.047 & $1.94 \pm 0.03$	& $2.58 \pm 0.18$	&13.6 \\
RGB J0710+591 &	0.124 & $1.53 \pm 0.12$ & $2.56 \pm 0.22$	& 8.3 \\
H 1426+428 &	0.129 & $1.32 \pm 0.12$ & $\sim \negthinspace 2.5$ & $\sim \negthinspace 9$ \\
1ES 0806+524 & 0.138 & $1.94 \pm 0.06$ & $3.6 \pm 1.0$ & 12.0 \\
1ES 0229+200 & 0.139 & $1.36 \pm 0.25$ & $2.58 \pm 0.14$ & 8.8 \\
RX J0648+1516 &	0.179 & $1.89 \pm 0.10$ & $4.40 \pm 0.80$ &14.0 \\
1ES 1218+304 & 0.182 & $1.71 \pm 0.07$	& $3.17 \pm 0.06$ & 8.0 \\
RBS 0413 & 0.190 & $1.55 \pm 0.11$ & $3.24 \pm 0.84$ & 8.9 \\
1ES 0414+009 & 0.287 & $1.98 \pm 0.16$	& $3.72 \pm 0.53$ & 6.1 \\
1ES 0502+675 & $\sim \negthinspace 0.341$? & $1.49 \pm 0.07$ & $3.92 \pm 0.35$ & 7.1? \\
PG 1553+113	& $0.4 \negthinspace \lesssim \negthinspace z \negthinspace \lesssim \negthinspace  0.5$ & $1.67 \pm 0.02$ & $4.37 \pm 0.09$ & 5.4 - 6.8 \\
VER J0521+211 & -- &	$1.93 \pm 0.03$ & $3.65 \pm 0.25$ & -- \\
\hline
\end{tabular}
\label{tab:hbls}
\end{center}
\end{table*}

\section{Science Motivations for HBL Observations}
\label{sec:science}
The science pursued through gamma-ray observations of HBLs is diverse -- ranging from the study of particle acceleration mechanisms at work in the collimated relativistic outflows of these sources, constraining/measuring diffuse cosmological radiation fields and intergalactic magnetic fields through the absorption and secondary emission of primary gamma-rays, to placing limits on source redshifts that remain elusive despite attempts at direct measurement.  These topics are each briefly discussed in the following subsections.

\subsection{Multiwavelength \& Variability Studies}
To fully understand the physical mechanisms at work in the jets of HBLs, it is critical to observe them from radio wavelengths up through gamma-ray energies. Doing so allows for correlation studies between X-ray and gamma-ray emission, searches for orphan gamma-ray flare events, the placement of constraints on emission models for different subclasses of AGN, and the general study of particle acceleration mechanisms in some of the most extreme environments in the Universe. VERITAS coordinates with a number of optical, X-ray, and gamma-ray instruments to maximize the science return for these types of investigations. 

Multiwavelength campaigns also serve the purpose of monitoring source behavior over extended periods of time.  This is key to establishing variability timescales which, in turn, are critical for determining the size of the emitting region as well as testing emission models involving (for example) line-of sight cosmic ray interactions \citep{Essey:2010} and gamma-ray-induced electromagnetic cascades \citep{Dermer:2011}. 

\begin{figure*}[t]
\centering
\includegraphics[width=155mm]{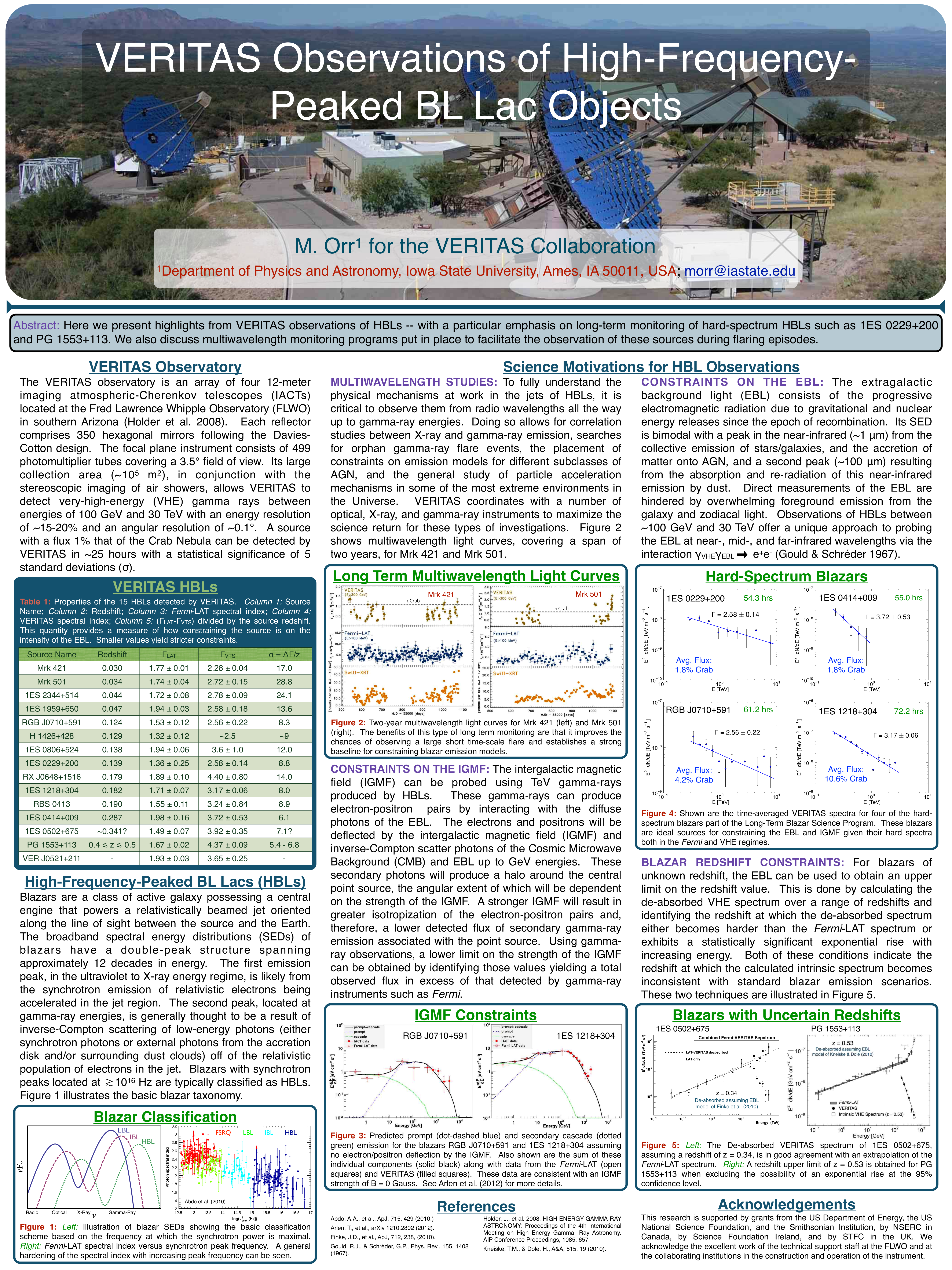}
\caption{Two-year multiwavelength light curves for Mrk 421 (left) and Mrk 501 (right).  The benefits of this type of long-term monitoring are that it improves the chances of observing a large, short time-scale flare and establishes a strong baseline for constraining blazar emission models.}
\label{fig:lightcurves}
\end{figure*}

Figure \ref{fig:lightcurves} shows multiwavelength light curves, for the two HBLs Mrk 421 and Mrk 501, covering a span of approximately two years from the fall of 2010 to the summer of 2012.

\subsection{Constraints on the Intergalactic Magnetic Field (IGMF)}
The intergalactic magnetic field (IGMF) can be probed using TeV gamma-rays produced by HBLs.  These primary gamma-rays can produce electron-positron  pairs by interacting with the diffuse photons of the EBL \citep{Huan:2011}.  The electrons and positrons will be deflected by the intergalactic magnetic field (IGMF) and inverse-Compton scatter photons of the cosmic microwave background (CMB) and EBL up to GeV energies.  These secondary photons will produce a halo around the central point source, the angular extent of which will be dependent on the strength of the IGMF.  A stronger IGMF will result in greater isotropization of the electron-positron pairs and, therefore, a lower detected flux of secondary gamma-ray emission associated with the point source.  Using gamma-ray observations, a lower limit on the strength of the IGMF can be obtained by identifying those values yielding a total observed flux in excess of that detected by gamma-ray instruments such as Fermi \citep{Taylor:2011,Arlen:2012}.

\begin{figure*}[t]
\centering
\includegraphics[width=155mm]{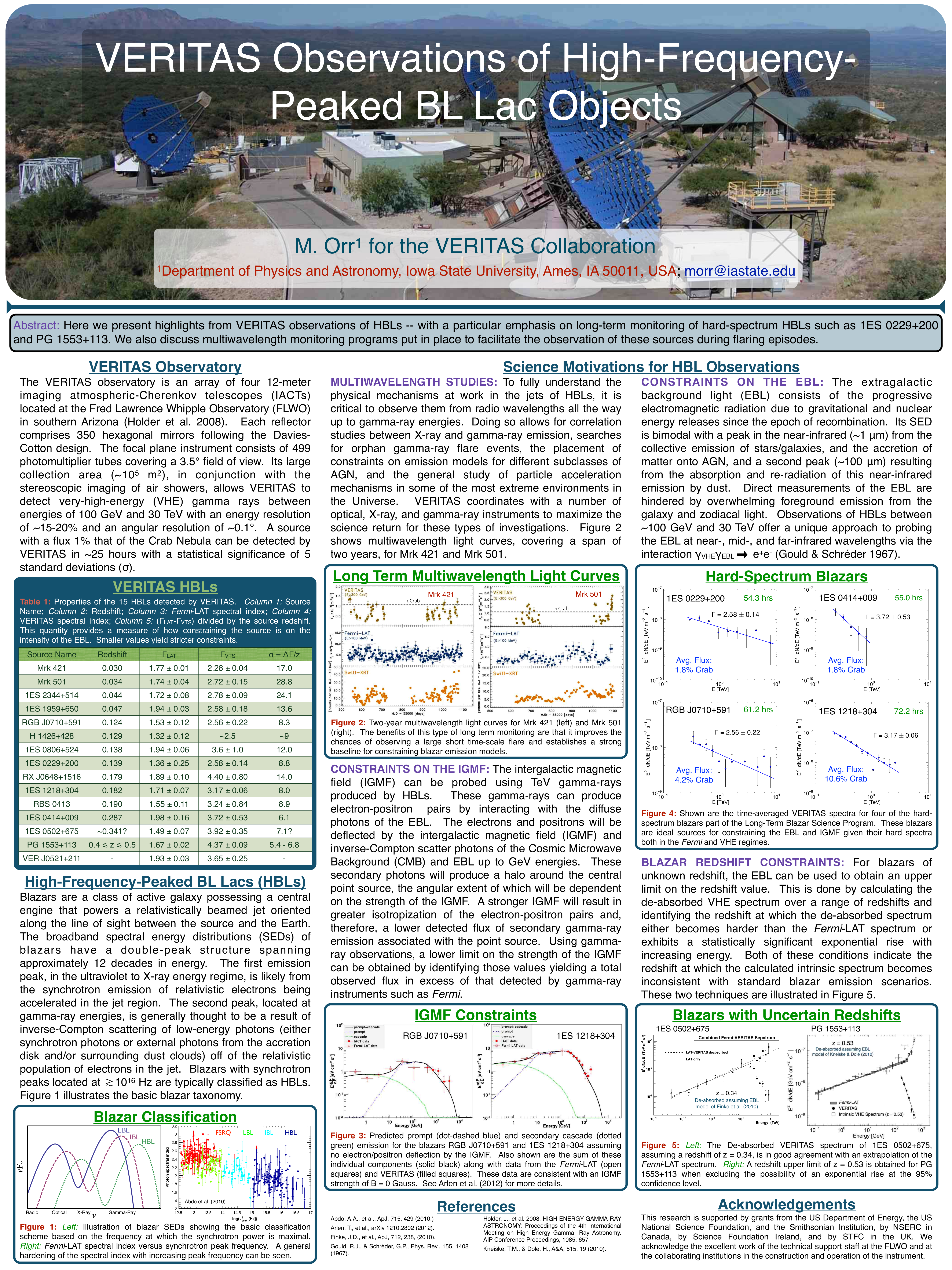}
\caption{Predicted prompt (dot-dashed blue) and secondary cascade (dotted green) emission for the blazars RGB J0710+591 and 1ES 1218+304 assuming no electron/positron deflection by the IGMF.  Also shown are the sum of these individual components (solid black) along with data from the Fermi-LAT (open squares) and VERITAS (filled squares).  These data are consistent with an IGMF strength of B = 0 Gauss.  See \citet{Arlen:2012} for more details.}
\label{fig:igmf}
\end{figure*}

Figure \ref{fig:igmf} shows the calculated primary and secondary emission components, assuming an IGMF strength of $B = 0\,$Gauss, for the HBLs RGB J0710+591 (left) and 1ES 1218+304 (right).  The \textit{Fermi} and VERITAS data shown for both sources are consistent with this assumed IGMF strength.  More details can be found in \citet{Arlen:2012}.

\subsection{Constraints on the Extragalactic Background Light (EBL)}
The extragalactic background light (EBL) consists of the accumulated electromagnetic radiation due to gravitational and nuclear energy releases since the epoch of recombination.  Its SED is bimodal with a peak in the near-infrared ($\sim \negthinspace 1\,\mu$m) from the collective emission of stars/galaxies, and the accretion of matter onto AGN, and a second peak ($\sim \negthinspace 100\,\mu$m) resulting from the absorption and re-radiation of this near-infrared emission by dust.  Direct measurements of the EBL are hindered by overwhelming foreground emission from the galaxy and zodiacal light.  Observations of HBLs between $\sim \negthinspace 100\,$GeV and 30 TeV offer a unique approach to probing the EBL at near-, mid-, and far-infrared wavelengths via the interaction $\gamma_\mathrm{VHE} \gamma_\mathrm{EBL} \rightarrow e^+ e^-$ \citep{Gould:1967}. 

Hard-spectrum blazars are best suited for EBL studies in this energy range when they have moderate to large redshifts and their hard observed gamma-ray spectra. The advantages associated with these characteristics are two-fold. First, one of the main effects of EBL absorption is that of a softening with respect to the intrinsic emission spectrum. If a source has a large redshift and hard observed spectrum, this places strong limits on the amount of EBL absorption that can be present without invoking an unrealistically (and un-physically) hard intrinsic spectrum in the very-high-energy (VHE) regime. Second, a hard observed spectrum means that the source in question is more readily detected at multi-TeV energies. This is critical in searching for a second form of EBL absorption signature - a spectral break at $\sim \negthinspace 1\,$TeV. These two EBL absorption signatures can be used in conjunction with one another to place strong constraints on the EBL \citep{Orr:2011}.

\begin{figure*}[p]
\centering
\includegraphics[width=155mm]{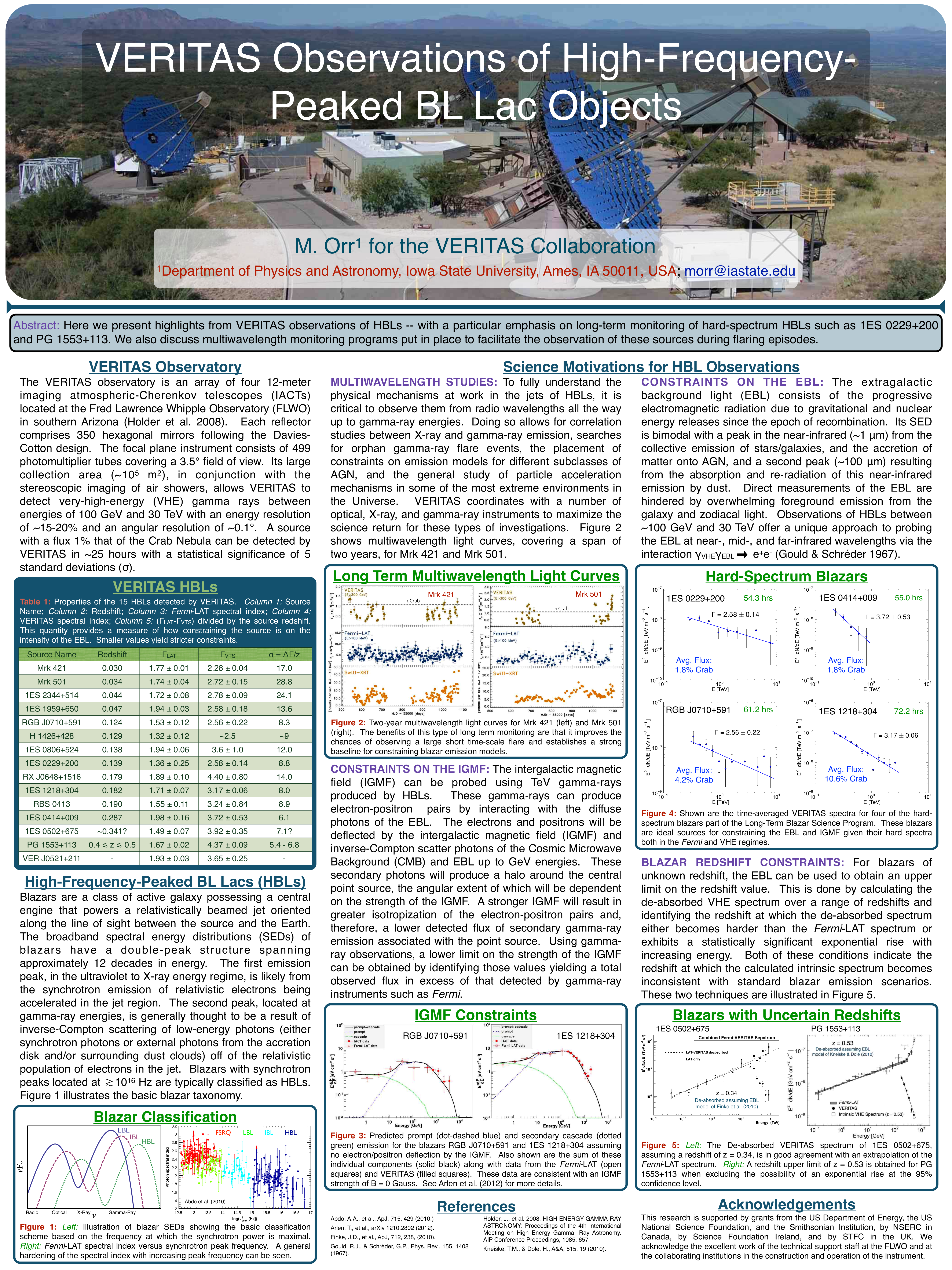}
\caption{Shown are the time-averaged VERITAS spectra for four of the hard-spectrum blazars that are part of the Long-Term Blazar Science Program.  These blazars are ideal sources for constraining the EBL and IGMF given their hard spectra both in the Fermi and VHE regimes.} 
\label{fig:spectra}
\end{figure*}

The VERITAS long-term blazar science and observing program includes the acquisition of deep observations on several hard-spectrum HBLs -- one of the goals behind these campaigns being improved constraints, or the potential measurement, of the EBL.  The spectra of four of these sources are shown in Figure \ref{fig:spectra}.

\subsection{Blazar Redshift Constraints}
\begin{figure*}[t]
\centering
\includegraphics[width=155mm]{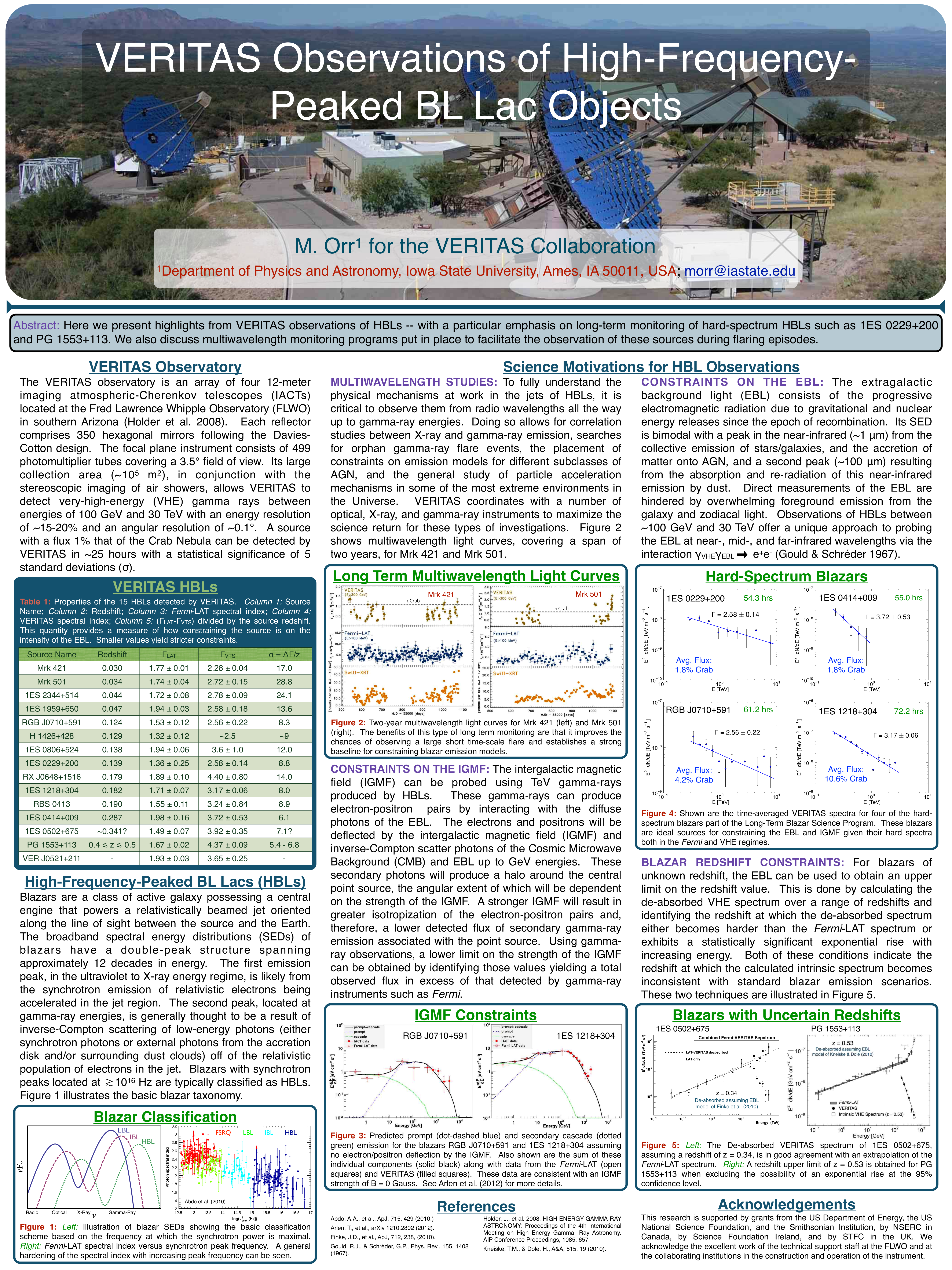}
\caption{\textit{Left:} The de-absorbed VERITAS spectrum of 1ES 0502+675, assuming a redshift of $z = 0.34$, is in good agreement with an extrapolation of the Fermi-LAT spectrum.  The de-absorbed spectrum was calcualted using the EBL model of \citet{Finke:2010}.  \textit{Right:} A redshift upper limit of $z = 0.53$ is obtained for PG 1553+113 when excluding the possibility of an exponential rise at the 95\% confidence level.  The EBL model used was that if \citet{Kneiske:2010}.} 
\label{fig:redshiftLimits}
\end{figure*}

For blazars of unknown redshift, the EBL can be used to obtain an upper limit on the redshift value.  This is done by calculating the de-absorbed VHE spectrum over a range of redshifts and identifying the redshift at which the de-absorbed spectrum either becomes harder than the Fermi-LAT spectrum or exhibits a statistically significant exponential rise with increasing energy.  Both of these conditions indicate the redshift at which the calculated intrinsic spectrum becomes inconsistent with standard blazar emission scenarios.  These two techniques are illustrated in Figure \ref{fig:redshiftLimits} using the VERITAS spectra of 1ES 0502+675 and PG 1553+113.  These data and analyses will be discussed in greater detail in upcoming dedicated publications.

\section{Conclusion}
\label{sec:conclusion}
The observation and study of HBLs are an important component of the VERITAS science program.  With continually deepening exposures on these objects, VERITAS will acquire the high-precision spectra necessary for performing detailed studies of the emission mechanisms at work in blazars, obtaining improved constraints on the EBL and potentially measuring its intensity, and placing the most stringent constraints on the strength of the IGMF to date.  The VERITAS collaboration is actively working with the multiwavelength community to ensure the greatest possible science return from these observations.

\begin{acknowledgments}
This research is supported by grants from the US Department of Energy, the US National Science Foundation, and the Smithsonian Institution, by NSERC in Canada, by Science Foundation Ireland, and by STFC in the UK. We acknowledge the excellent work of the technical support staff at the FLWO and at the collaborating institutions in the construction and operation of the instrument.
\end{acknowledgments}

\bigskip 

\end{document}